\documentclass{PHYEAUTH}
\usepackage{graphicx}
\usepackage{amsmath}
\usepackage{amssymb}
\def\esp{\Delta E_{\rm SP}}
\def\ef{E_{\rm F}}

\begin{document}

\begin{frontmatter}

\title{Geometric Suppression of Single-Particle Energy Spacings in Quantum
Antidots}

\author[address1]{L.~C.~Bassett\corauthref{cor}},
\author[address2]{C.~P.~Michael},
\author[address1]{C.~J.~B.~Ford},
\author[address1]{M.~Kataoka},
\author[address1]{C.~H.~W.~Barnes},
\author[address3]{M.~Y.~Simmons}
and
\author[address1]{D.~A.~Ritchie}

\address[address1]{Cavendish Laboratory, J.~J.~Thompson Avenue, Cambridge CB3 OHE, UK}

\address[address2]{Department of Applied Physics, California Institute of Technology, Pasadena, CA 91125, USA}

\address[address3]{School of Physics, University of New South Wales, Sydney 2052, Australia}

\corauth[cor]{ Corresponding author. E-mail: lcb36@cam.ac.uk}

\begin{abstract}
Quantum Antidot (AD) structures have remarkable properties in the
integer quantum Hall regime, exhibiting Coulomb-blockade charging
and the Kondo effect despite their open geometry.  In some regimes a
simple single-particle (SP) model suffices to describe experimental
observations while in others interaction effects are clearly
important, although exactly how and why interactions emerge is
unclear.  We present a combination of experimental data and the
results of new calculations concerning SP orbital states which show
how the observed suppression of the energy spacing between states
can be explained through a full consideration of the AD potential,
without requiring any effects due to electron interactions such as
the formation of compressible regions composed of multiple states,
which may occur at higher magnetic fields.  A full understanding of
the regimes in which these effects occur is important for the design
of devices to coherently manipulate electrons in edge states using
AD resonances.
\end{abstract}

\end{frontmatter}

Several exciting developments in the field of mesoscopic physics
have recently emerged from the study of coherent electronic devices
\cite{Neder2007}. By utilising coherent edge states in quantum Hall
systems, devices can be constructed which coherently manipulate
electron spin, charge, and phase, offering important prospects for
the study of quantum information in the solid state.  Quantum
antidots in particular offer spin-selective coherent control of
these states, making them potentially important in future
applications \cite{Zozoulenko2004}. An antidot (AD) is simply a
potential hill in a two-dimensional electron system, which in a
perpendicular magnetic field ($\gtrsim 0.3$~T) gains a set of
quasibound states quantised by the Aharonov-Bohm (AB) effect. When
placed at the centre of a constriction, these states couple to
extended edge channels, and their structure may then be investigated
through the scattering processes which determine the conductance. In
the simplest model, these noninteracting single-particle (SP) states
adjust with changing magnetic field in order to maintain their
enclosed flux, producing a set of magnetoconductance oscillations as
they pass through the Fermi energy ($\ef$) in the leads. Many
details of AD behaviour may be understood within this SP picture,
particularly at low fields ($\lesssim 3$~T) \cite{Mace1995}.  Other
effects, such as Coulomb Blockade \cite{Kataoka1999} and the Kondo
Effect \cite{Kataoka2002} require a self-consistent model including
electron interactions.\footnote{For a recent review see
\cite{Sim2007}.}  There has been some debate over the precise nature
of these effects \cite{Karakurt2001,Kataoka2004,Goldman2004},
particularly concerning the presence or absence of compressible
regions (CRs) composed of multiple partially-occupied SP states
within a few $k_{\rm B} T$ of $\ef$. Certainly, in the limit of
large AD size and high field, CRs are expected to form in order to
minimise the Coulomb energy associated with abrupt changes in
electron density \cite{Chklovskii1992}, but these may be suppressed
in other regimes by the AB quantisation and exchange effects. Recent
numerical calculations accounting for both Coulomb and
spin-dependent interactions confirm this expectation
\cite{Ihnatsenka2006a}; for an AD of radius 200~nm in the integer
quantum Hall regime with filling factor $\nu=2$, CRs are
significantly quenched for fields below $\approx4$~T, and then a CR
only forms for the outer spin state.

\begin{figure}
\includegraphics[width=3truein]{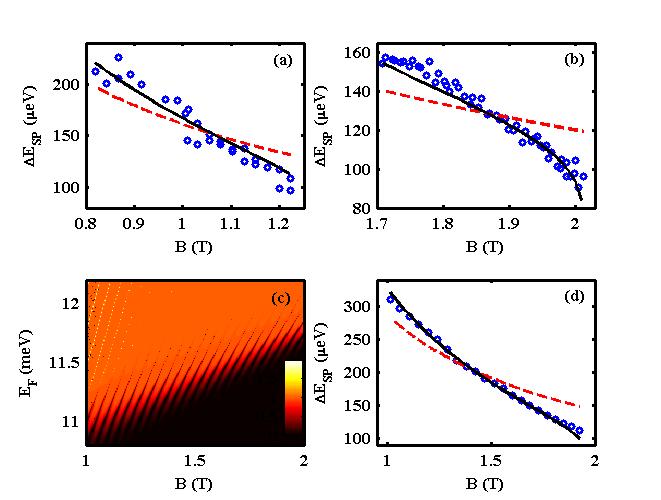}
\caption{Top panels:  Single-particle energy $\esp$ (circles)
extracted from DC-bias measurements in different ranges of magnetic
field (with different gate voltages). A $1/B$ fit (dashed curve)
fails to match the data while our model (solid curve) predicts the
reduction of $\esp$ at higher fields. Bottom panels: Conductance (c)
in units of $2e^2/h$ as a function of $B$ and $\ef$, calculated from
the full (noninteracting) Green's function computed using an
iterative procedure \cite{MacKinnon1985}, and the corresponding
energy spacing (d), calculated at $\ef=11.7$~mV.  As in the
experimental data, our model (solid curve) accounts for the
discrepancy from the $1/B$ dependence (dashed curve).}
\label{fig:fits}
\end{figure}

We present a series of measurements designed to explore the
evolution of SP states with magnetic field.  Our device\footnote{
Complete details may be found in \cite{Michael2006}.} consists of an
AD gate (200~nm in diameter) centered in a 1-$\mu$m channel,
contacted by a metal gate isolated from the split gate by an
insulating layer of crosslinked polymer, allowing us to control the
voltage on the AD independently.  As a function of $B$ and
source-drain bias, the conductance forms a complicated pattern of
resonances not unlike that observed in quantum dots.  From these
data, we can extract the various transition energies of the system
--- the Coulomb energy $E_{\mathrm C}$, the Zeeman spin-splitting
$E_{\mathrm Z}$, and the SP energy $\esp$. As was noted in
\cite{Michael2006}, at higher fields we observe a clear suppression
of $\esp$ below the expected $1/B$ dependence for circular AD
states, and it was then suggested that this could reflect the
formation of CRs.  We have investigated this further, however, and
find that this suppression occurs systematically on the high-$B$ end
of the $\nu=2$ plateau, coinciding with the transition of AB
resonances from transmission peaks to reflection valleys.  We have
also seen this effect at relatively low fields $\lesssim2$~T (see
Fig.~\ref{fig:fits}) where CRs are less likely to form.  We
therefore propose an alternative explanation for this effect, namely
that the presence of the split-gate spreads out the AB states in the
constrictions, leading to a suppression of $\esp$ for the lowest SP
states in a Landau level.

\begin{figure}
\includegraphics[width=3truein]{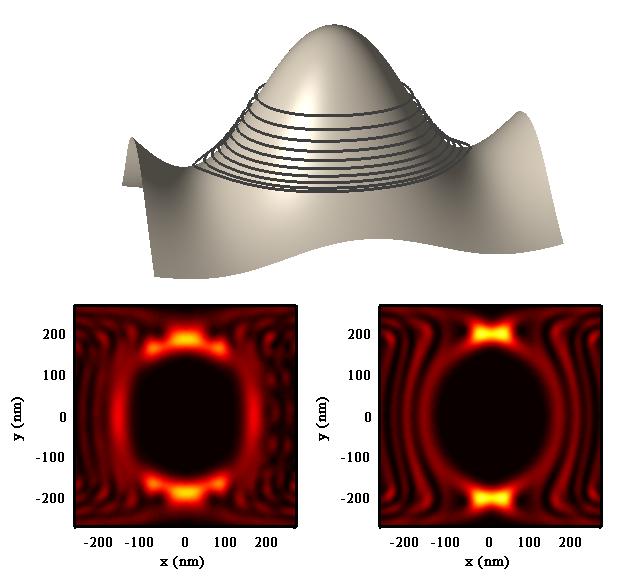}
\caption{Top:  Potential created by an AD in a constriction
(computed as in \cite{Davies1995}), with SP contours calculated
according to Eq.~(\ref{eq:esp}).  Note the ``bulging'' of the
contours into the constrictions, which results in a much-reduced
energy spacing. Bottom:  Local density of states calculated from the
noninteracting Green's function for the AD shown above.  Note that
the transmission resonance (left panel) at low field
($\approx\!1$~T) is more circularly symmetric than the reflection
resonance (right panel) at higher field ($\approx\!1.5$~T).
\label{fig:GF}}
\end{figure}

It is straightforward to show that the asymmetry introduced by the
constrictions leads to a suppression of $\esp$.  The spatial
separation of SP states is determined by the additional flux $h/e$
enclosed by each successive orbit, and the ``bulging'' of states
lower in the constrictions accounts for an increased fraction of the
area.  Since the states are contours of the AD potential and the
slope far away from the constrictions is nearly constant, the energy
spacing between these states is reduced (see Fig.~\ref{fig:GF}). For
an AD potential $U(r,\theta)$ which varies on a scale much larger
than the separation between states, we can calculate
\begin{equation}
\esp = -\frac{h}{eB}\left[ \int_0^{2\pi}\left(\frac{dU}{dr}
\right)_{(\mathcal{C},\theta)}^{-1}\mathcal{C}(\theta)d\theta^{-1}\right],
\label{eq:esp}
\end{equation}
where $\mathcal{C}(\theta)$ is the contour at the Fermi energy
defined by $U_{\rm eff}(\mathcal{C},\theta) = \ef$, using the
effective potential $U(r,\theta) + E_{\rm cyc} + E_{\rm Z}$, where
$E_{\rm cyc}$ and $E_{\rm Z}$ are the cyclotron and Zeeman energies
for states in the constrictions. For a circularly symmetric
potential this gives the $1/B$ dependence mentioned
above,\footnote{The $B$-dependence of $\mathcal{C}(\theta)$ is quite
weak.} but it is clear that a reduction in the slope of the
potential over any region of the contour results in a suppressed
$\esp$.  It is worth noting that the magnetic flux does not solely
determine the location of AD states, since the nonuniform AD
potential contributes an additional phase factor to the
wavefunctions beyond the magnetic AB phase,\footnote{This also means
that AD states do not enclose integer multiples of the flux quantum
$h/e$ as commonly assumed, but rather are pushed outwards by the AD
potential.} but for a potential that varies sufficiently slowly, the
dominant contribution to the difference between adjacent states
arises from the change in magnetic flux, and so a model considering
only differences in area is appropriate.

Using the bare electrostatic potential \cite{Davies1995} resulting
from the gates on our device, we can use Eqn.~\ref{eq:esp} to
calculate $\esp$ as a function of $B$ to compare with the data.  In
Fig.~\ref{fig:fits} (a and b) we show the results of this
calculation for two data sets with different experimental
parameters, and a comparative best-fit curve $\propto 1/B$.  The
model itself has no free parameters; the potential is completely
determined by the arrangement of gates, the measured AB period
(related to the AD area by $A\Delta B = h/e$), and $\esp$ at low
field, and we can estimate $E_{\rm F}$ from the field at which the
$\nu = 2$ state is depopulated in the channel.  We have included no
effects of tunnelling between SP states and the leads in this
calculation, which results in an artificial drop to zero as the
saddle point reaches $\ef$ and closed orbits no longer exist.  For
additional comparison with this essentially classical model, we have
calculated the full (noninteracting) Green's function for an AD $+$
split-gate geometry using an iterative procedure
\cite{MacKinnon1985}. Fig.~\ref{fig:GF} shows the calculated local
density of states for an AD device at a reflection resonance, and
Fig.~\ref{fig:fits}c shows the calculated conductance as a function
of $B$ and $E_{\rm F}$.  From the conductance we can calculate
$\esp$, as shown in Fig.~\ref{fig:fits}d, and we find a nearly
identical suppression at higher fields which is well matched by our
geometric model. Since this calculation includes no electron
interaction effects, we can be sure that CRs are not required to
produce this behaviour.

We therefore conclude that the observed suppression of $\esp$ is a
simple result of the potential profile in our experimental geometry,
rather than a signature of a reorganisation of states into a CR.
Although we know that interaction effects become essential for an
understanding of AD resonances at high fields, this study
demonstrates the capacity of the SP model to explain relatively
complicated features of the excitation spectrum of ADs in the
low-field regime.  An understanding of this structure is critical in
the design of AD-based applications which seek to utilise processes
in a specific regime.  By choosing AD sizes and fields
appropriately, it may be possible to construct devices which use ADs
in different regimes (even on a single chip) to manipulate spin or
charge for different purposes.  These avenues of research remain
largely unexplored, and may hold many exciting advances in the study
of quantum-coherent electronics.

We acknowledge financial support from the Marshall Aid Commemoration
Commission and the NSF Graduate Research Fellowship, and
computational resources provided by CamGrid.

\bibliographystyle{elsart-num}

\end{document}